# Two-phase Interactions between Propagating Shock Waves and Evaporating Water Droplets

Zhiwei Huang*, Huangwei Zhang
Department of Mechanical Engineering, National University of Singapore, 9 Engineering Drive 1, Singapore 117576, Republic of Singapore
*Corresponding author email: mpehuan@nus.edu.sg

## Abstract
One-dimensional numerical simulations based on hybrid Eulerian-Lagrangian method are performed to study the interactions between propagating shocks and dispersed evaporating water droplets. Two-way coupling for exchanges of mass, momentum, energy and vapour species is adopted for the dilute two-phase gas-droplet flows. Interphase interactions and droplet breakup dynamics are investigated with initial droplet diameters of 30, 50, 70 and 90 μm under an incident shock wave Mach number of 1.3. Novel two-phase flow phenomena are observed when droplet breakup occurs. First, droplets near the two-phase contact surface show obvious dispersed distribution because of the reflected pressure wave that propagates in the reverse direction of the leading shock. The reflected pressure wave grows stronger for larger droplets. Second, spatial oscillations of the gas phase pressure, droplet quantities (e.g., diameter and net force) and two-phase interactions (e.g., mass, momentum, and energy exchange), are observed in the post-shock region when droplet breakup occurs, which are caused by shock / droplet interactions. Third, the spatial distribution of droplets (i.e., number density, volume fraction) also shows strong oscillation in the post-shock region when droplet breakup occurs, which is caused by the oscillating force exerted on the droplets.

## Introduction
Shock / droplet interactions are frequently encountered in engineering practice. Examples are the high-speed vehicle flying through the cloud (droplet diameters of $d_{d,0}$ = 1-100 μm) or rain ($d_{d,0}$ = 100-10,000 μm) [1] and the liquid jet of hydrocarbon fuels ($d_{d,0}$ = 10-200 μm) in scramjet engines [2]. If the aerodynamic force is strong enough, a large droplet can be deformed and broken into finer droplets. This phenomenon, known as the secondary atomization, has been extensively studied over decades [3–7]. A common example is the aerodynamic breakup of water droplets (WDs) impinging with shocks. The shock is attenuated in both strength and propagation speed because of the WD breakup and the interphase exchanges of mass, momentum and energy, whereas the WDs are broken into smaller ones and are evaporated, accelerated and heated by the shock and post-shock gas. This is of particular significance for industrial fire and explosion mitigation as WDs can extract energy from the shock or blast waves (heat absorption effects [8]) and slow down the chemical reactions (dilution effects [9]), mainly attributes to their large heat capacity, easy and cheap acquisition and environmentally safety.

In this study, numerical simulations are performed with hybrid Eulerian-Lagrangian approach to investigate the two-phase interactions between propagating shock waves and evaporating water droplets proceed with breakup. Evolutions of droplet diameter and two-phase exchange terms are provided. The specific flow phenomena resulted from droplet breakup are discussed in detail. This differs from the previous studies, which have been focused on the morphologies of shock structure [1,10–13] and/or WD breakup dynamics [14–16].

**Numerical method**

Full details about the governing equations for both gas and droplet phases have been provided in our recent work [17]. The compressible two-phase flow solver, *RYrhoCentralFoam* [18], also has been introduced therein. It is developed from the *rhoCentralFoam* solver [19] in OpenFOAM 5.0 package, and is validated against a range of benchmark cases including the Sod's shock tube problem, forward-facing step, supersonic jet and shock-vortex interaction [19–21]. Furthermore, the *RYrhoCentralFoam* solver and implementation of droplet submodels (e.g. evaporation and multi-species diffusion) also have been validated and verified with a series of canonical tests against analytical solution and/or experimental data in our recent work [22,23].

The implicit second-order Crank-Nicolson scheme is used for temporal discretization. The semi-discrete Kurganov, Noelle and Petrova (KNP) scheme [24] with a Minmod flux limiter [25] is used for convective fluxes, whereas the second-order central differencing scheme [26] is applied for diffusive fluxes. The Lagrangian equations for the droplet phase are integrated with a first-order implicit Euler method. The CFL number of the gas phase equations is 0.02, which corresponds to the physical time step of about $10^{-8}$ s.

**Physical problem**

One-dimensional (1D) computational domain is used in this work as in our previous study [17]. It is not intended to capture the detailed morphology of either the shock or WDs as such topics have been stressed in the previous studies with two- or three-dimensional configurations [1,10,12,14–16,27,28]. The 1D simplification also has been widely used for particle [29–31] or droplet [11,13,32,33] laden flows with shocks.

Figure 1(a) shows a schematic of the 1D domain ($x$ = 0-1.6 m) and the initial distribution of WDs. A right propagating shock enters the domain at $x$ = 0 and $t$ = 0. WDs are monodispersed and uniformly distributed in the droplet-laden section ($x$ = 0.8-1.6 m). The length between the left end and left boundary of the two-phase region (0.8 m) is long enough to ensure that the reflected pressure wave ($M_{rp}$ in Fig. 1b) does not arrive at the left end in the interested time (from when the incident shock enters the two-phase region to that when the leading shock reaches the right end).

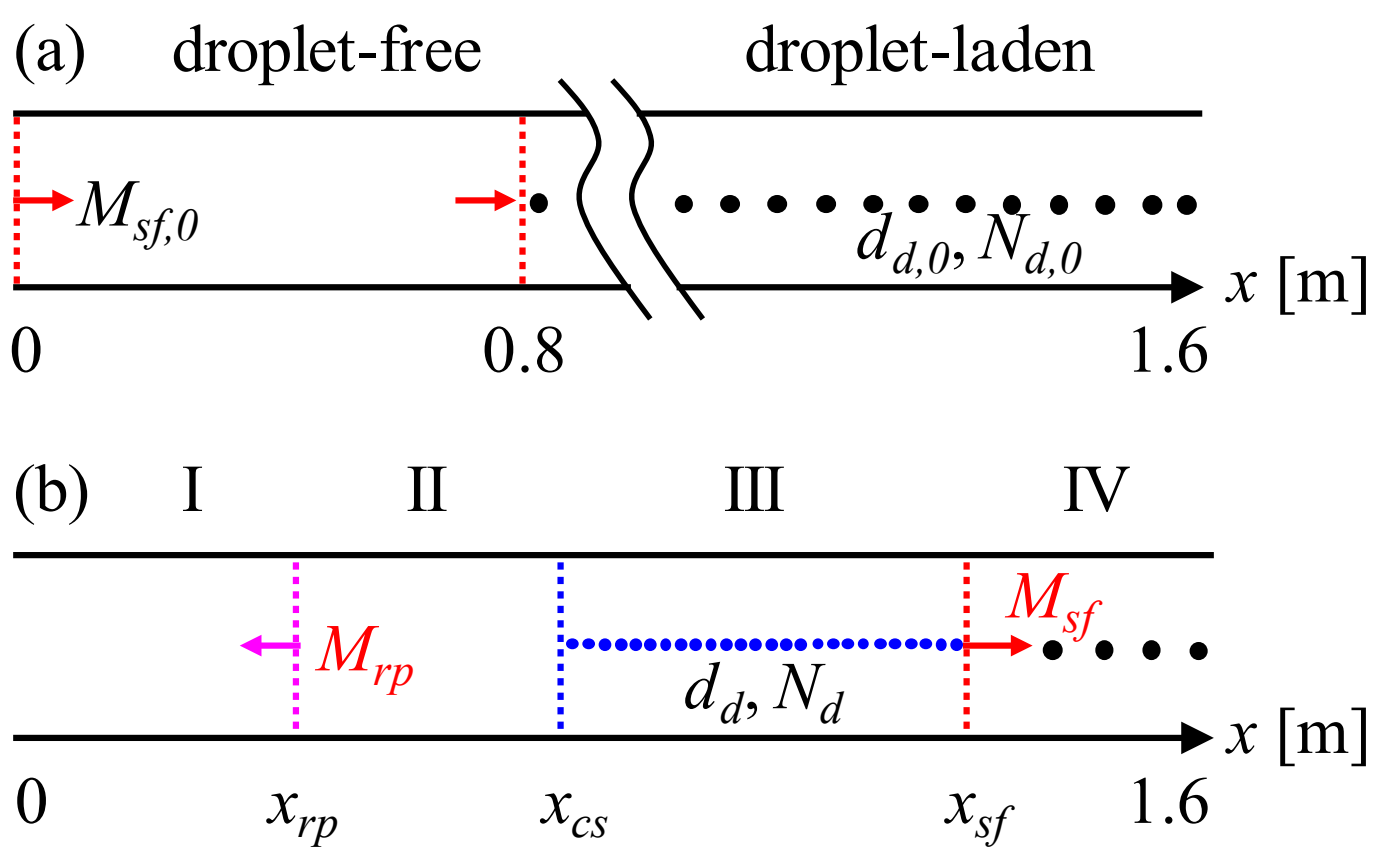

**Figure 1.** Schematic of shock propagation in water clouds: (a) before and (b) after the shock enters the two-phase region. $M_{sf,0}$ / $M_{sf}$ are the incident / instantaneous shock Mach numbers, $d_{d,0}$ / $N_{d,0}$ and $d_d$ / $N_d$ are the initial and instantaneous diameter / number density of water droplets. Circles: water droplets.

Figure 1(b) shows an instantaneous scenario after the shock propagates into the two-phase region. Three characteristic fronts are observed in our simulations, i.e., the leading shock, two-phase contact surface (PCS, interface of the purely gaseous and droplet-laden regions) and reflected pressure wave (RPW). Their instantaneous locations are respectively denoted as $x_{sf}$, $x_{cs}$ and $x_{rp}$. Four zones of interest are resulted from the characteristic fronts. Zone I (0 - $x_{rp}$) is compressed by the incident shock, II ($x_{rp}$ - $x_{cs}$) is compressed by the RPW and is droplet-free, III ($x_{cs}$ - $x_{sf}$) is the droplet breakup zone and IV ($x_{sf}$ - 0.16 m) is the unperturbed (for both gas and WDs) pre-shock zone.

**Simulation conditions**

The initial carrier gas is $O_2$ / $N_2$ mixture with mass fraction of 0.233 / 0.767, respectively. In the pre-shock region, the initial gas temperature and pressure are 293 K and 10 kPa, respectively. The considered incident shock Mach number and droplet number density are $M_{sf,0}$ = 1.3 and $N_{d,0}$ = 2.5 × $10^7$ /$m^3$, respectively. The initial density, heat capacity and temperature of WDs are respectively 996.3 kg/$m^3$, 4189.9 J/kg/K and 293 K. Four different initial droplet diameters, i.e., $d_{d,0}$ = 30, 50, 70 and 90 μm are investigated. The breakup model used in the present work is the classical TAB model [34].

The droplet volume fraction ($V_{fd}$) is generally below 0.1% and diluted medium is assumed [35]. Hence, the interactions between droplets (e.g. collision and coalescence) are not considered. Furthermore, the internal viscous force is small compared with surface tension force for WDs and hence is omitted. Both the carrier gas and WDs are quiescent at $t$ = 0.

Dirichlet conditions are applied at the left boundary of the computational domain in Fig. 1. Non-reflective condition is applied at the right boundary. The domain is discretized with a uniform cell size of 1 mm. This grid resolution also has been used and validated with finer meshes (cell sizes of 0.3 and 0.1 mm) in our recent work [17], which has shown that both the gas and droplet behaviours are not sensitive to the Eulerian mesh resolution.

**Results and Discussion**

Figure 2 shows the evolutions of the droplet diameter ($d_d$) in the $x$-$t$ diagram. At the right side of lines $L_{sf,1}$ - $L_{sf,4}$, the droplets are undisturbed and they evaporate slowly. Meanwhile, at the left side of lines $L_{de,1}$ - $L_{de,4}$, the region is droplet-free. It is seen in Figs. 2(a) and 2(b) that there is absolutely no breakup of droplets for $d_{d,0}$ = 30 and 50 μm. Droplet diameters only decrease slightly (about 0.2 μm from their initial values) because of evaporation behind the shock. This value is significantly smaller than that with smaller initial diameters in our recent work [17]. For example, for droplets with $d_{d,0}$ = 5 um under an incident shock of $M_{sf,0}$ = 1.35, droplet diameter can be decreased to about 1 μm near the two-phase contact surface, although the post-shock gas conditions of the two incident shocks (i.e., $M_{sf,0}$ = 1.35 in our previous study [17] and $M_{sf,0}$ = 1.3 in this work) are close. This can be explained through the $d^2$-law, i.e., to evaporate the same mass rate of vapour, larger droplet decreases slower in diameter because of its larger surface area.

In Figs. 2(c) and 2(d), droplet breakup occurs immediately after the shock passage, i.e., $d_d$ is discontinuous across the shock fronts (lines $L_{sf,3}$ - $L_{sf,4}$). After breakup, $d_d$ decreases slowly and continuously in the region from the shock front to near the PCS (lines $L_{de,3}$ - $L_{de,4}$) because of evaporation. Noticeable dispersion of droplet distribution is observed next to the PCS for this two cases, which is caused by the left-propagating RPW (shown in Fig. 3). The final droplet diameter because of breakup is about 4 μm, which indicates low sensitivity to the initial droplet diameter.

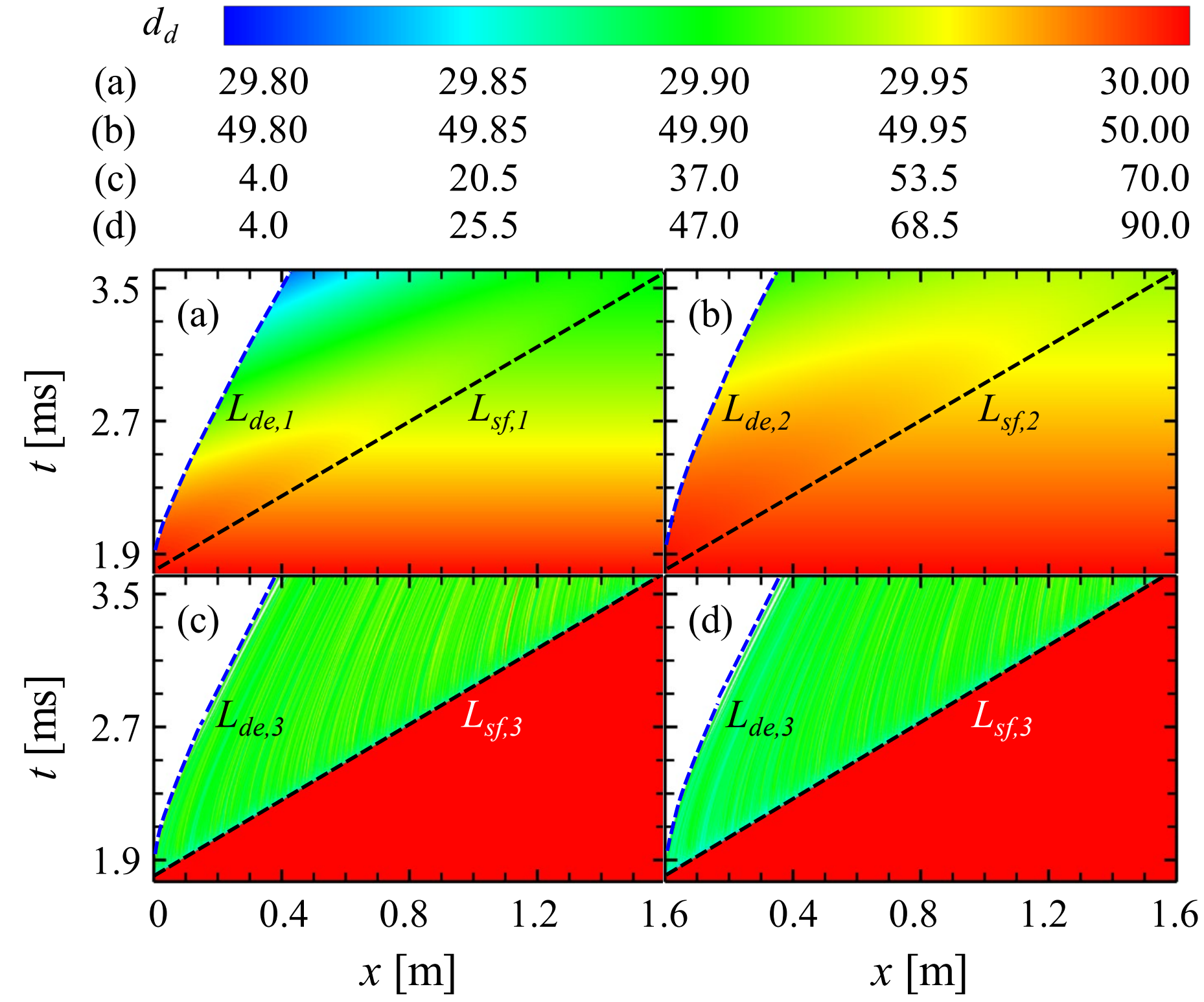

**Figure 2.** *x-t* diagram of droplet diameters (in μm) with initial diameters of (a) 30, (b) 50, (c) 70 and (d) 90 μm. $L_{sf,1}$ - $L_{sf,4}$: leading shock, $L_{de,1}$ - $L_{de,4}$: two-phase contact surface.

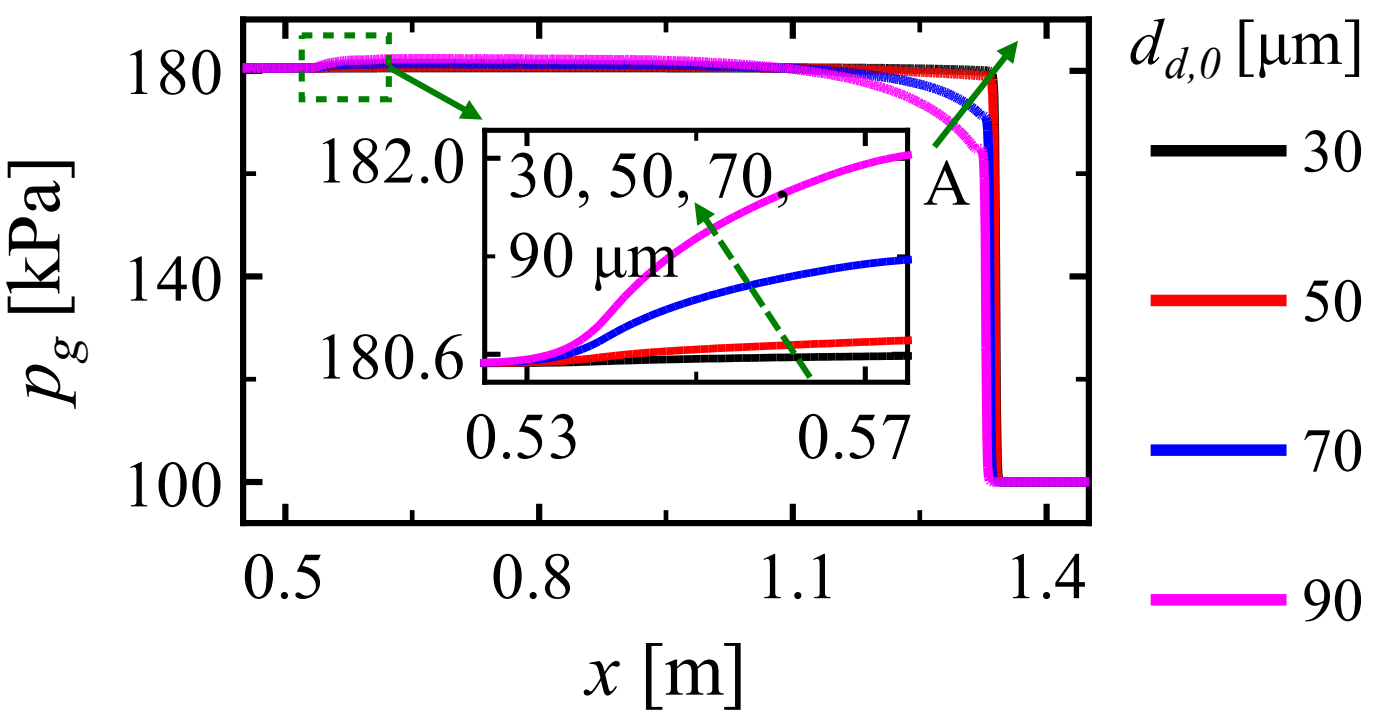

**Figure 3.** Comparisons of the gas phase pressure for different initial droplet diameters at $t$ = 3 ms.

Figure 3 shows the distribution of gas phase pressure ($p_g$) at $t$ = 3.0 ms for the above four cases. In the inset, it is seen that a left-propagating RPW is formed because of the consecutive interactions between the right-propagating shock and WDs. Such phenomenon is also observed in other studies, e.g., in Refs. [11,31]. However, the RPW is not significant for small WDs (e.g., $d_{d,0}$ = 5-20 μm in our previous study [17]) as it is too weak to be observed. In Fig. 3, it is obvious that the reflected pressure increases rapidly with droplet diameter. This left-propagating pressure wave decreases the velocity of the right-propagating droplets near the PCS. Hence, those droplets near the PCS gradually fall behind other droplets that are less affected by the RPW. Furthermore, for $d_{d,0}$ = 70 and 90 μm, the shock fronts (arrow A) have been smoothed at this instant. It is obvious that larger droplet volume fraction ($V_{sf}$, which is proportional to the cubic of diameter) has stronger attenuation effects on the shock strength [17]. However, the shock propagation speed is little affected by $V_{sf}$ as the shock fronts almost reach at the same location for the four cases. This is because the overall droplet volume fraction (lower than $10^{-4}$ for all cases in Fig. 5 shown later) is much lower than that can lead to obvious shock attenuation in our recent work [17]. For example, significant shock attenuation only occurs when $V_{sf} > 10^{-3}$ for $M_{sf,0} = 1.35$ (i.e., Fig. 14 in Ref. [17]).

Figure 4 shows the evolutions of the interphase momentum exchange rate in the *x*-*t* diagram for the above-mentioned four cases. For the breakup-free cases (i.e., $d_{d,0}$ = 30 and 50 μm), $S_M$ is spatially continuous in the post-shock region as those for small droplets in our previous work [17].The highest $S_M$ follows right behind the leading shock and decays towards the PCSs. Note that $S_M \approx 0$ in the droplet-laden region means almost no interphase velocity difference, and hence, all droplets share the same velocity to the local gas. In addition, the momentum relaxation zone (i.e., where $S_M$ is significant) is almost constant (however is extended for larger droplets, e.g., from $d_{d,0}$ = 30 μm to 50 μm) in width with respect to time. This leads to almost uniform distribution of the droplet volume fraction after some distance of the leading shock in Figs. 5(a) and 5(b).

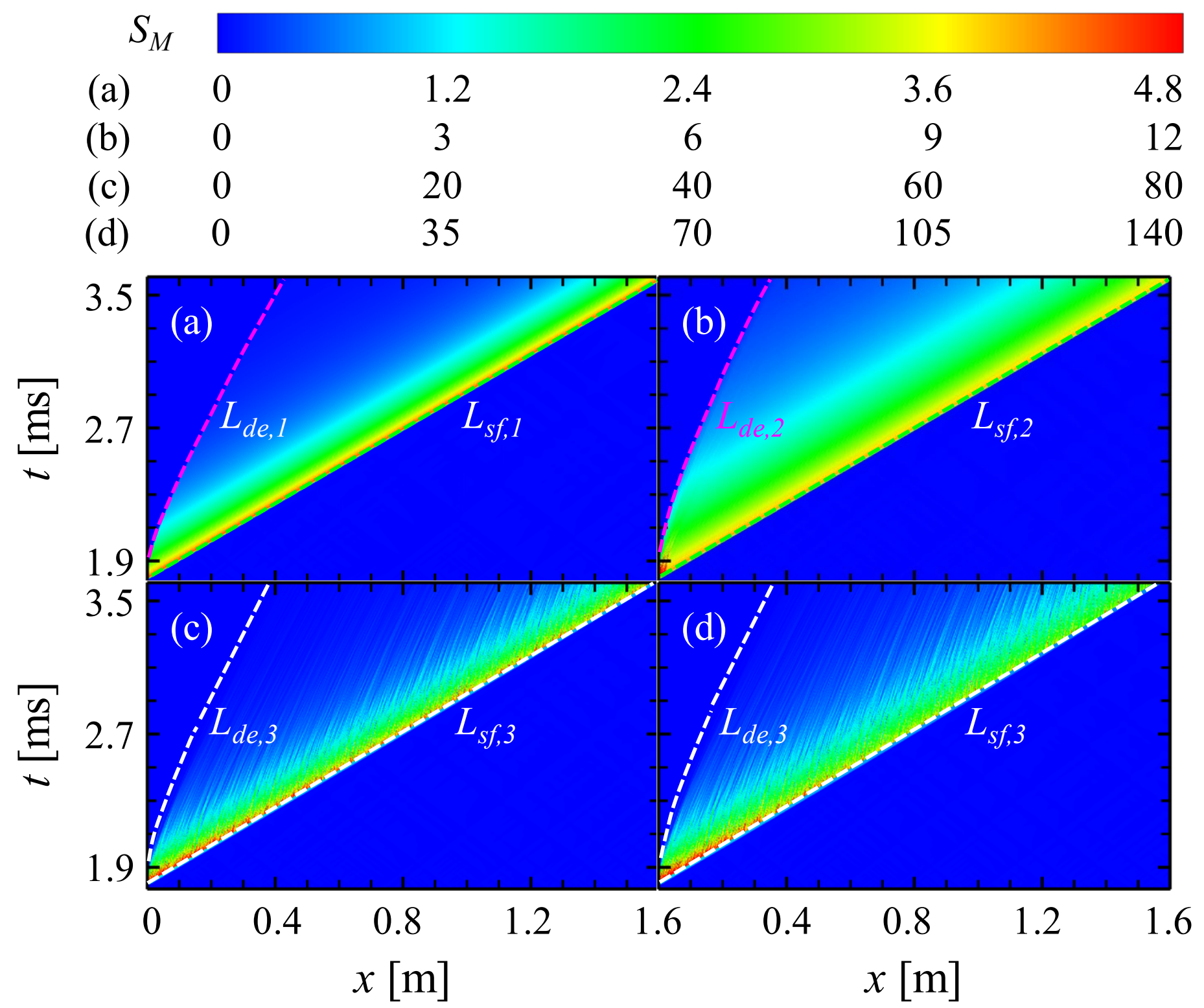

**Figure 4.** *x*-*t* diagram of interphase momentum transfer rate (in kN/m$^3$) with initial droplet diameters of (a) 30, (b) 50, (c) 70 and (d) 90 μm. The description for dashed lines is the same as in Fig. 2.

For the droplet breakup cases ($d_{d,0}$ = 70 and 90 μm), the interphase momentum exchange is discontinuous and shows strong spatial oscillation in the post-shock region. This is because of the spatially discretized distribution of WDs (indicated by $V_{sf}$) with respect to axial location in Figs 5(c) and 5(d). The reason is that larger droplets are broken into smaller ones, which have better performance in response to the variation of gas flow velocity. Note that the droplet momentum response timescale is proportional to the square of its diameter [35]. On the other hand, the RPW becomes significant for large droplets (e.g., $d_{d,0}$ = 70 and 90 μm in Fig. 3). A local RPW is generated for each individual droplet when it interacts with the leading shock, which leads to an accumulation of various local and non-uniform reflected pressure waves. The reflected pressure wave shown in Fig. 3 is only the final stable one. This also applies for other droplet quantities besides volume fraction, e.g., number density (not shown for length limit). Hence, the interphase exchange terms, e.g., momentum, mass and energy (the latter two are also not shown here), show significant oscillation in space.

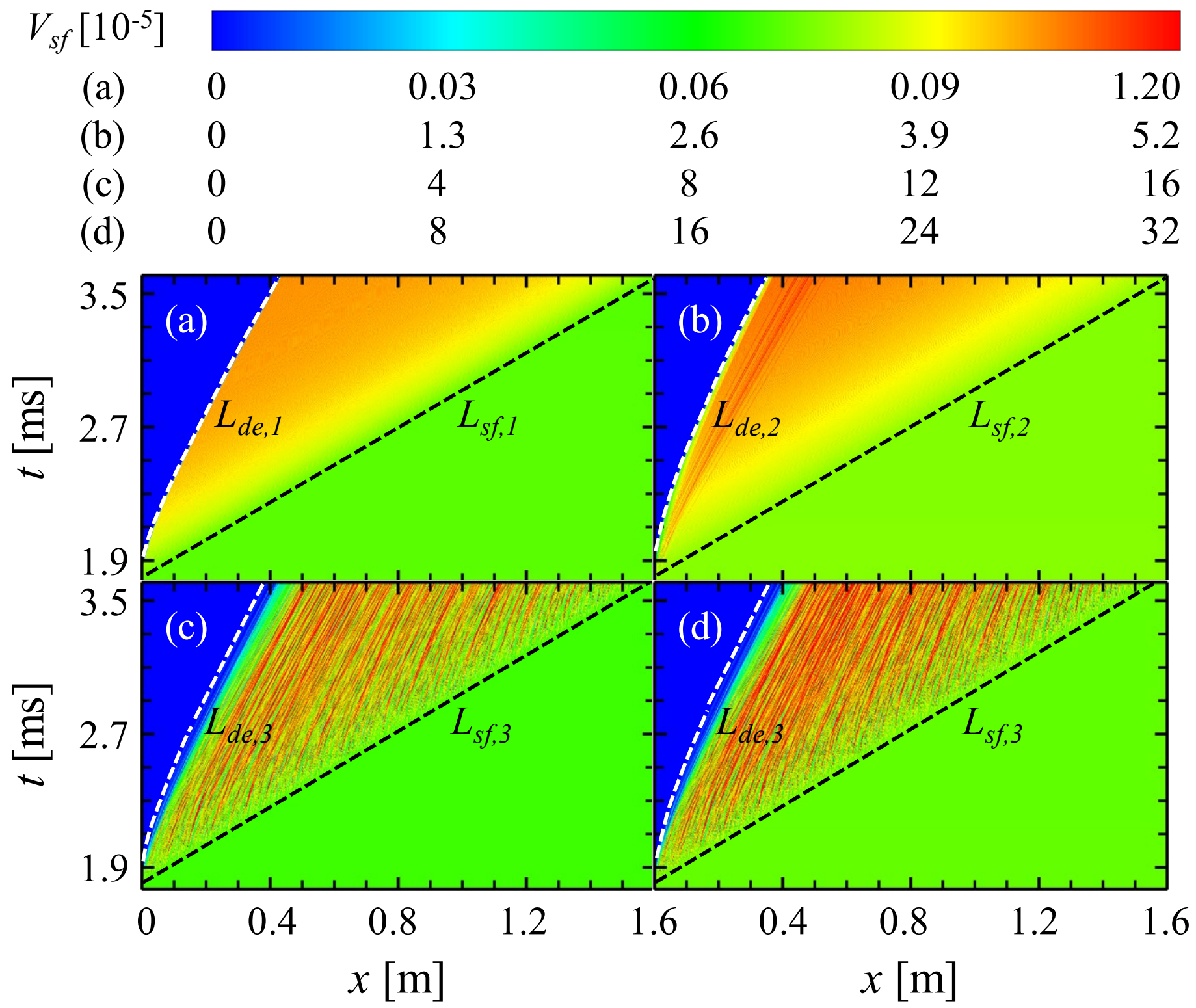

**Figure 5.** *x-t* diagram of droplet volume fraction with initial droplet diameters of (a) 30, (b) 50, (c) 70 and (d) 90 μm. The description for dashed lines is the same as in Fig. 2.

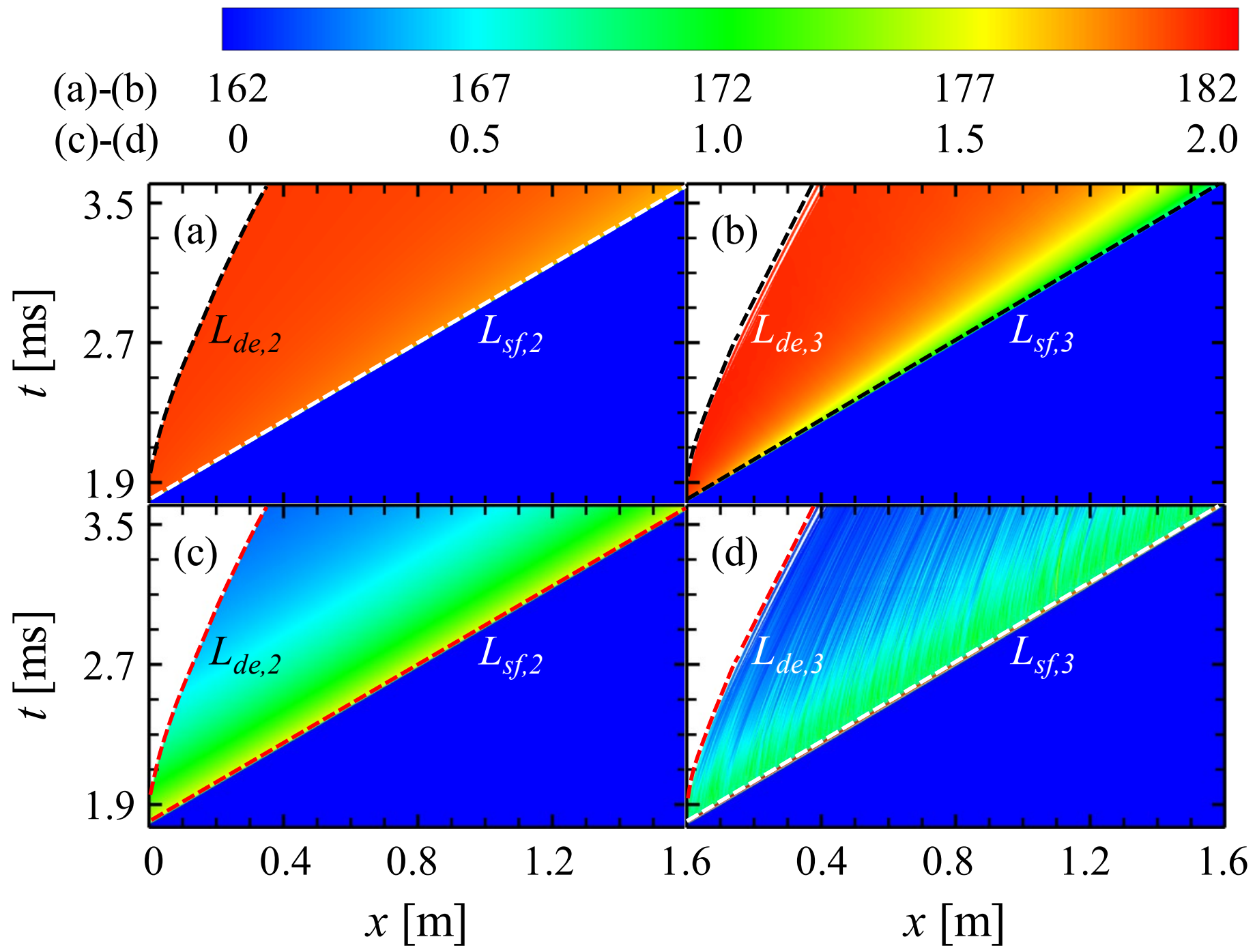

**Figure 6.** *x-t* diagram of (a)-(b) gas phase pressure (in kPa) and (c)-(d) net force (in mN) exerted on droplets. (a) and (c) for $d_{d,0}$ = 50 μm, (b) and (d) for $d_{d,0}$ = 70 μm. The description for dashed lines is the same as in Fig. 2.

To further demonstrate the local pressure waves (lead to pressure non-uniformity in the post-shock zone) and droplet location oscillation, Fig. 6 shows the evolutions of gas phase pressure (interpolated to the droplet position) and net force (the sum of drag force and pressure gradient force, $F_n$) exerted on the droplets for $d_{d,0}$ = 50 and 70 μm exemplarily. The non-uniformity of $p_g$ behind the shock front is weak for $d_{d,0}$ = 50 μm in Fig. 6(a), whereas it becomes significant for $d_{d,0}$ = 70 μm in Fig. 6(b). Furthermore, the overall tendency of $p_g$ increases from the shock front to the two-phase contact surface because of the accumulation of $p_g$ in space. That is to say, for the regions farther from the shock front, the more local RPWs have been accumulated and the pressure is higher. Specifically, $p_g$ increases continuously in space for $d_{d,0}$ = 50 μm, whereas its jump is more obvious for $d_{d,0}$ = 70 μm. The non-uniform $p_g$ in Fig. 6(b) is mainly resulted from the local RPWs because of the shock / droplet interactions.

It is seen that $F_n$ is smooth for $d_{d,0}$ = 50 μm in Fig. 6(c), but $F_n$ spatially oscillates for $d_{d,0}$ = 70 μm in Fig. 6(d). This is the major reason for the non-uniform distribution of droplets observed in Figs. 4-5. The pressure non-uniformity because of shock / droplet interaction and the spatial oscillation of droplet distribution in the post-shock region are not observed for small droplets ($d_{d,0}$ = 5-20 μm) in our previous study [17] as well as the relatively large droplets with $d_{d,0}$ = 30 and 50 μm in this study. Furthermore, these phenomena also exist for other incident shock Mach numbers (e.g., $M_{sf,0}$ = 1.5, 1.7 and 1.9) based on our numerical experimentation, as long as droplet breakup occurs.

**Conclusions**

One-dimensional numerical simulations are conducted to study the intersections between the propagating shocks and dilute evaporating water droplets. Droplet breakup and exchanges in mass, momentum, energy and vapour species between the carrier gas and water droplets are respectively considered through the TAB breakup model and two-way coupling of the Eulerian-Lagrangian approach. Three characteristic fronts are observed when the shock travels in the two-phase gas-droplet medium: a leading shock that propagates right with the fastest velocity, a contact surface that separates the droplet-laden region from the droplet-free region, and a reflected pressure wave that propagates in a reverse direction against the leading shock.

The gas phase quantities (e.g., pressure, temperature, velocity and vapour mass fraction) show strong oscillation spatially because of the oscillating two-phase interactions when droplet breakup occurs. The droplet quantities (e.g., diameter, volume fraction and number density) and two-phase exchange terms (e.g., mass, momentum and energy transfer rates) also show significant spatial oscillations, which are caused by the local shock / droplet interactions. A local reflected pressure wave is generated for each individual droplet when impinges with the leading shock, which accumulates in the post-shock region, from the leading shock front to the two-phase contact surface. This leads to obvious non-uniform net force exerted on the droplets, which finally leads to the highly oscillating distribution of water droplets in space.

**Acknowledgments**

This work was financially supported by the Singapore MOE Tier 1 grant (R-265-000-653-114). The simulations are performed with the computational resources from National Supercomputing Center Singapore (NSCC, https://www.nscc.sg/).